\documentstyle[prl,aps,epsf,epsfig]{revtex}

\def\be{\begin{equation}}
\def\ee{\end{equation}}
\def\bea{\begin{eqnarray}}
\def\eea{\end{eqnarray}}
\def\bd{\begin{displaymath}}
\def\ed{\end{displaymath}}

\newdimen\normalarrayskip              % skip between lines
\newdimen\minarrayskip                 % minimal skip between lines
\normalarrayskip\baselineskip
\minarrayskip\jot
\newif\ifold             \oldtrue            
\def\arraymode{\ifold\relax\else\displaystyle\fi} % mode of array entries
\def\@arrayskip{\ifold\baselineskip\z@\lineskip\z@
     \else
     \baselineskip\minarrayskip\lineskip2\minarrayskip\fi}
\def\@arrayclassz{\ifcase \@lastchclass \@acolampacol \or
\@ampacol \or \or \or \@addamp \or
   \@acolampacol \or \@firstampfalse \@acol \fi
\edef\@preamble{\@preamble
  \ifcase \@chnum
     \hfil$\relax\arraymode\@sharp$\hfil
     \or $\relax\arraymode\@sharp$\hfil
     \or \hfil$\relax\arraymode\@sharp$\fi}}
\def\@array[#1]#2{\setbox\@arstrutbox=\hbox{\vrule
     height\arraystretch \ht\strutbox
     depth\arraystretch \dp\strutbox
     width\z@}\@mkpream{#2}\edef\@preamble{\halign \noexpand\@halignto
\bgroup \tabskip\z@ \@arstrut \@preamble \tabskip\z@ \cr}%
\let\@startpbox\@@startpbox \let\@endpbox\@@endpbox
  \if #1t\vtop \else \if#1b\vbox \else \vcenter \fi\fi
  \bgroup \let\par\relax
  \let\@sharp##\let\protect\relax
  \@arrayskip\@preamble}
\makeatother

\newlength{\extraspace}
\newlength{\extraspaces}
\begin{document}
\draft
\twocolumn[\hsize\textwidth\columnwidth\hsize\csname
@twocolumnfalse\endcsname

\title{
\begin{flushright}
{\large \tt CERN-TH/99-137}
\end{flushright}
%\begin{center}
\rule{17.3cm}{0.2mm}
{\Large \bf  
 The Pinch Technique Approach to \\the Physics of Unstable Particles } 
\rule{17.3cm}{0.2mm} 
%\end{center}
}

\author{{\large \bf Joannis Papavassiliou} \\
\vspace*{0.2 cm}
{\large  CERN Theory Division} \\
{\large CH-1211 Geneva 23, Switzerland} \\
\vspace*{0.45 cm}
{\normalsize \tt Contribution to
the 1998 Corfu Summer Institute  on \\
Elementary Particle Physics (JHEP proceedings)
}}

\maketitle
.

\begin{abstract}
 {\normalsize {\large \bf Abstract:}
The consistent description of unstable particles within the
framework of perturbative gauge field theories necessitates 
the definition and resummation of off-shell Green's functions,
which must respect several crucial physical requirements.
We present the solution to this problem at one-loop, using  
the pinch technique.
\vspace*{0.4 cm} }
\end{abstract}
]

\begin{narrowtext}

\section
{ Resonances and the need for resummation} 

The physics of unstable particles   in general \cite{THUN}
 and the computation  of
resonant transition amplitudes in particular 
\cite{PHUN}
has attracted significant
attention  in  recent  years, because  it   is both phenomenologically
relevant   and  theoretically  challenging. 
In what follows we will formulate the problem in simple terms, before
we proceed to its actual solution.

The mathematical expressions for computing
transition amplitudes 
are ill-defined
in the vicinity  of resonances, because  the tree-level propagator
of the particle mediating the interaction, i.e. $\Delta= (s-M^2)^{-1}$, 
becomes singular as the center-of-mass energy $\sqrt{s}\sim M$. 
The standard way for regulating this physical kinematic singularity
is to use a  Breit-Wigner  type of
propagator, which essentially amounts to the replacement
$(s-M^2)^{-1} \to (s-M^2+iM\Gamma)^{-1}$, where $\Gamma$ is the
width of the unstable (resonating) particle. The field-theoretic
mechanism which enables this replacement is the Dyson resummation of
the (one-loop) self-energy $\Pi(s)$
of the  unstable particle, which leads to the substitution  
$(s-M^2)^{-1} \to [s-M^2+\Pi(s)]^{-1}$; the running width of the particle
is then defined as $M\Gamma(s) =\Im m \Pi(s)$, whereas the
usual (on-shell) width is simply its value at $s=M^2$.

It is well-known that, to any finite order, 
the conventional perturbative expansion
gives rise to expressions for physical amplitudes 
which are endowed with 
crucial 
properties. For example, the amplitudes are independent of the
gauge-fixing parameter (GFP) chosen to quantize the theory, they 
are gauge-invariant (in the sense of current conservation), they 
are unitary (in the sense of probability conservation), and well behaved
at high energies. The above properties are however not always encoded
into the individual Green's functions which are the building blocks
of the aforementioned perturbative expansion; indeed,
the simple fact that  Green's functions
depend in general explicitly on the GFP,
indicates that they are void of any physical meaning. Evidently,
when going from unphysical 
Green's functions to physical amplitudes subtle field-theoretical 
mechanisms are at work, which implement highly non-trivial
cancellations among the various Green's functions appearing at a given
order.

The happy state of affairs described above is guaranteed 
within the framework of the
conventional perturbative expansion, provided that 
one works at a given fixed order.
It is relatively easy to realize however that 
the Breit-Wigner
procedure is in fact equivalent to a reorganization
of the perturbative series; indeed, resumming the
self-energy $\Pi$ amounts to removing a particular piece from each
order of the perturbative expansion, since from all the Feynman 
graphs contributing to a given order $n$ we only pick the part that contains   
$n$ self-energy bubbles $\Pi$, and then take $n \to \infty$.
However, given that a non-trivial
cancellation involving the unphysical Green's function
is generally taking place at any given order of the 
conventional 
perturbative expansion,
the removal of one of them from 
each order may or may not distort 
those cancellations. To put it differently, if  $\Pi$ contains
unphysical contributions (which would eventually cancel 
against other
pieces within a given order) resumming it naively would mean
that these unphysical contributions have also undergone 
infinite summation (they now appear in the denominator of the propagator
$\Delta$). In order to remove them one has to add the remaining 
perturbative pieces to an infinite order, clearly an impossible
task. Thus, if the resummed $\Pi$ happened to contain such 
unphysical terms, one would finally arrive 
at predictions for  the amplitude close to the resonance
which would be plagued with  
unphysical
artifacts. 
It turns out that, while in scalar field theories and Abelian gauge
theories  $\Pi$ does not contain such unphysical contributions,
this seizes to be true in the case of   non-Abelian gauge  
theories.

The most obvious signal revealing that 
the conventionally defined non-Abelian self-energies
are not good candidates for resummation  
comes from   the
simple calculational fact that  the  bosonic radiative corrections  to
the  self-energies of  vector  ($\gamma$, $W$, $Z$)  or scalar (Higgs)
bosons  induce a non-trivial  dependence on the
GFP used to define the tree-level  bosonic propagators appearing in
the   quantum  loops.   This is  to    be contrasted to  the fermionic
radiative corrections, which, even in the context of non-Abelian gauge
theories behave as in quantum  electrodynamics (QED), {\em i.e.}, they
are   GFP    independent.    In addition,    formal    field-theoretic
considerations as well  as direct calculations  show that, contrary to
the  QED case, the non-Abelian  Green's functions do not satisfy their
naive, tree-level  Ward   identities (WI's),  after bosonic   one-loop
corrections  are  included.   A  careful   analysis  shows that   this
fundamental difference  between Abelian  and non-Abelian  theories has
far-reaching  consequences;   the     naive  generalization   of   the
Breit-Wigner method to   the latter case  gives rise  to Born-improved
amplitudes, which do  not faithfully capture the underlying  dynamics.
Most notably,  due  to  violation of   the  optical theorem,
unphysical thresholds and  artificial resonances appear, which distort
the line-shapes of the resonating particles.  
In addition, the high energy properties of
such amplitudes are altered, and  are in direct contradiction to the
equivalence theorem (ET) \cite{EqTh}.

In order to address these issues, 
 a new approach
to resonant transition amplitudes has been developed over the past 
few years \cite{PP2,PPHiggs}, 
which
is based on the 
the pinch technique (PT)~\cite{PT,PT2};
the latter is a diagrammatic method whose main 
thrust is  to
exploit  the symmetries built  into  physical amplitudes in
order   to construct 
off-shell sub-amplitudes which are kinematically akin to conventional
Green's 
functions, but, unlike the latter, are also endowed with several
crucial properties: 
(i) they are independent of  the GFP; 
(ii) they satisfy naive (ghost-free) tree-level  Ward identities (WI's) 
instead of the usual Slavnov-Taylor  identities;     
(iii) they display    physical
thresholds  only \cite{PP2}; 
(iv)  they  satisfy  individually the optical   
and
equivalence theorems \cite{PP2,PRW,PPHiggs}; 
(v) they are analytic functions of the kinematic
variables; 
(vi)  the   effective two-point functions  constructed  are
universal (process-independent) \cite{NJW0},  
Dyson-resummable \cite{PP2,NJW2},  and do not  shift
the  position of the  gauge-independent complex  pole \cite{PP2}.  
The crucial novelty introduced by the PT is that
the resummation of graphs must take place only
{\em after} the amplitude of interest has been cast
via the PT algorithm into
manifestly physical sub-amplitudes, with distinct
kinematic properties, order by order in perturbation theory.
Put in the language employed earlier, the PT 
ensures that all unphysical contributions contained inside $\Pi$ have
been identified and properly discarder, before $\Pi$ undergoes 
resummation. 
It is important to emphasize that the only ingredient needed for
constructing the PT effective Green's functions is the full exploitation
of elementary Ward-identities (EWI), (which are a direct consequence
of the BRS \cite{BRS} symmetry of the theory)
and the proper use of the unitarity
and analyticity of the $S$-matrix.
In what follows we will describe the 
method in detail.

\section{ The Pinch Technique rearrangement of the amplitude.}

 Within the PT framework, the transition
amplitude $T(s,t,m_i)$ of a $2\to 2$ process,
can be decomposed as
\begin{equation}
\label{TPT}
T(s,t,m_i)\ =\ \widehat{T}_1(s)\ +\ \widehat{T}_2(s,m_i)\ +\
\widehat{T}_3(s,t,m_i),
\end{equation}
in terms of three individually g.i.\ quantities:
a propagator-like part ($\widehat{T}_1$), a vertex-like piece
($\widehat{T}_2$),
and a part containing box graphs ($\widehat{T}_3$). The important observation
is that vertex and box graphs contain in general
pieces, which are kinematically akin to self-energy graphs
of the transition amplitude.
The PT is a systematic way of extracting such pieces and
appending them to the conventional self-energy graphs.
In the same way, effective gauge invariant
vertices may be constructed, if
after subtracting from the conventional vertices the
propagator-like pinch parts we add the vertex-like pieces coming from
boxes. The remaining purely box-like contributions are then
also gauge invariant. The way to identify the pieces which are
to be reassigned, all one has to do is to resort to the fundamental
PT cancellation, which is in turn a direct consequence of the
elementary Ward identities of the theory. This cancellation 
is depicted in Fig. for the process $e^{+}e^{-}\to W^{+}W^{-}$, and
will be studied in detail in the next sections.

The PT rearrangement of the amplitude has far-reaching consequences.
Perhaps the best way to appreciate them is to study 
the
close   connection which exists
between  gauge invariance and  unitarity; the latter 
is best established by
looking at the two  sides of the equation for  the optical theorem.  
The  optical theorem for a
given process $\langle a|T|a \rangle$ is
\begin{equation}
  \label{OT}
\Im m \langle  a|T|a\rangle \  =\ \frac{1}{2}\,
\sum_{f}\int  \langle  f|T|a\rangle   \langle f|T|a\rangle^{*}
\, ,
\end{equation}
where  the sum  $\sum_f$ should be  understood  to be  over the entire
phase space and spins of  all possible on-shell intermediate particles
$m$.   The RHS   of   Eq.\ (\ref{OT})  consists   of  the  product  of
GFP-inde-\\
pendent on   shell      amplitudes,   thus   enforcing     the
gauge-invariance of the imaginary part of the amplitude on the LHS. In
particular, even though the LHS contains unphysical particles, such as
ghosts   and  would-be Goldstone  bosons,  which   could  give rise to
unphysical    thresholds, Eq.(\ref{OT})   guarantees   that all   such
contributions    will  vanish.    In    general,   the  aforementioned
cancellation  takes place  after  contributions from  the propagator-,
vertex-, and box-diagrams   have   been combined.   There   are  field
theories however, such as  scalar theories, or  QED, which allow for a
stronger version of the equality given in Eq.\ (\ref{OT}): The optical
relationship holds  {\em  individually} for the  propagator-, vertex-,
and box-diagrams.  
In non-Abelian  gauge  theories however, the  afore-mentioned stronger
version of the optical theorem
does {\em not} hold in  general.  The reason is that
unlike     their scalar or    Abelian   counterparts, the conventional
self-energies, vertex and boxes  are {\em gauge dependent}. 

As  has been  demonstrated in a  series   of papers \cite{PP2,PRW}
however, a strong version of the optical theorem very analogous to that
depicted in Fig.2 
can be realized  in the context of non-Abelian  gauge theories  at one
loop, if the amplitudes are rearranged according to the PT algorithm.
Specifically, let us apply  the PT on  both sides of  Eq.\ (\ref{OT}):
The PT rearrangement of the tree-level cross sections appearing in the
RHS  gives  rise to new  process-independent (self-energy-like) parts,
which are  equal to the  imaginary part of the effective self-energies
obtained by  the application of the  PT on the one-loop expression for
the amplitude $\langle a|T|b\rangle$ on the LHS  .  The same result is
true for the vertex- and  box-like parts, defined  by the PT on either
side  of Eq.(\ref{OT}).     In other  words, effective  sub-amplitudes
obtained after  the   application  of  the  PT satisfy   the 
optical theorem
{\em
  individually}, {\em e.g.},
\begin{equation}
  \label{OTPT}
\Im m \Big( \langle  a|T|a\rangle_{\rm PT}^{j}\Big)\  =\ \frac{1}{2}\,
\sum_{f}\int  \Big(   \langle  f|T|a\rangle   \langle f|T|a\rangle^{*}
\Big)_{\rm PT}^{j}\, ,
\end{equation}
where the subscript  ``PT''  indicates that the PT   rearrangement has
been carried  out, and   the  index $j=S,V,B$,  distinguishes  between
effective self-energy, vertex, and boxes, respectively.

Turning to the question of how a 
resonant amplitude should be regulated, 
the strategy is now clear:
We begin from the RHS of the optical relation given in Eq.\ 
(\ref{OT}). The RHS involves on-shell physical processes, which
satisfy the EWIs. The full exploitation of those EWIs leads
unambiguously to a decomposition of the tree-level amplitude into
propagator-, vertex- and box-like structures. The propagator-like structure
corresponds to the imaginary part of the effective propagator under
construction. By imposing the additional requirement that the effective
propagator be an analytic function of $q^2$ one arrives at a 
dispersion relation, which,
up to renormalization-scheme choices, leads to a unique result for the real 
part. In the next three section we will study exactly how this strategy 
is implemented. 
First we will study the relevant field-theoretical
aspects in the case of QCD, which, even though does not allow for
(fundamental) resonances, captures most of the issues one needs to 
understand. Then  we will address the electroweak case,
and finally we will turn to the particularities of the Higgs-boson resonance.

\section{ The case of QCD}

Consider the forward scattering process $q {\bar q}\rightarrow q{\bar q}$,
shown in Figure 4.
 From
the optical theorem, we then have
\begin{equation}
\Im m \langle q\bar{q}|T|q\bar{q}\rangle\ =\ \frac{1}{2}\, 
\left( \frac{1}{2} \right)\, 
\int dX_{LIPS}\,  
|\langle q\bar{q}|T|gg\rangle|^2\, .
\label{OTgg}
\end{equation}
In Eq.\ (\ref{OTgg}), the statistical factor 1/2 in parentheses arises from
the fact that the final on-shell gluons should be considered as identical
particles in the total rate. The integration measure $dX_{LIPS}$ 
denotes the two-body Lorentz invariant phase-space.
 We now set ${\cal M}=\langle
q\bar{q}|T|q\bar{q}\rangle $ and ${\cal T}=\langle q\bar{q}|T|gg\rangle$, and
focus on the RHS of Eq.\ (\ref{OTgg}). 

Diagrammatically, the amplitude ${\cal
T}$ consists of two distinct parts: $t$ and $u$-channel graphs that contain an
internal quark propagator, ${{\cal T}_{t}}^{ab}_{\mu\nu}$, as shown in Figs.\
3(a) and 3(b), and an $s$-channel amplitude, ${{\cal T}_{s}}^{ab}_{\mu\nu}$,
which is given in Fig.\ 3(c). The subscript ``$s$'' and ``$t$'' refers to the
corresponding Mandelstam variables, {\em i.e.}\ $s=q^2=
(p_1+p_2)^2=(k_1+k_2)^2$, and $t=(p_1-k_1)^2=(p_2-k_2)^2$. 
Defining 
\begin{equation}
V_{\rho}^{c}\ =\ g\bar{v}(p_2)\, \frac{\lambda^c}{2}\gamma_{\rho}\, u(p_1)\, ,
\end{equation}
we have that
\begin{equation}
{\cal T}^{ab}_{\mu\nu}={{\cal T}_{s}}^{ab}_{\mu\nu} (\xi )+
{{\cal T}_{t}}^{ab}_{\mu\nu}\, ,
\label{DefT}
\end{equation}
with
\begin{eqnarray}
{{\cal T}_{s}}^{ab}_{\mu\nu}(\xi ) & =&
-gf^{abc}\, \Delta^{(\xi ),\rho\lambda}_0(q)
\Gamma_{\lambda\mu\nu}(q,-k_1,-k_2)\, V_{\rho}^{c}\, ,
\nonumber\\
{{\cal T}_{t}}^{ab}_{\mu\nu} &=& -ig^2\bar{v}(p_2)\Big( 
\, \frac{\lambda^b}{2}\gamma^{\nu}\, \frac{1}{\not\! p_1-\not\! k_1 - m}
\, \frac{\lambda^a}{2}\gamma^{\mu}\ +\ \nonumber\\
&& \frac{\lambda^a}{2}\gamma^{\mu}\,
\frac{1}{\not\! p_1-\not\! k_2-m}\, 
\gamma^{\nu}\frac{\lambda^b}{2}\, \Big)u(p_1)\, ,\qquad 
\label{Tt}
\end{eqnarray}
where
\bea
\Gamma_{\lambda\mu\nu}(q,-k_1,-k_2)\ &=&\ 
(k_1-k_2)_{\lambda}g_{\mu\nu}\, +\, (q+k_2)_{\mu}g_{\lambda\nu}\nonumber\\
&& -\, (q+k_1)_{\nu}g_{\lambda\nu}\, .
\label{3GV}
\eea
Notice that ${\cal T}_{s}$ depends explicitly on the GFP $\xi$, through the
tree-level gluon propagator $\Delta^{(\xi)}_{0\mu\nu}(q)$, whereas ${\cal
T}_{t}$ does not. The explicit expression of $\Delta^{(\xi )}_{0\mu\nu}(q)$
depends on the specific gauge fixing procedure chosen.  In addition, we define
the quantities ${\cal S}^{ab}$ and ${\cal R}^{ab}_{\mu}$ as follows:
\begin{eqnarray}
{\cal S}^{ab}&=&gf^{abc}\, \frac{k^\sigma_1}{q^2}\, 
V_{\sigma}^{c}\nonumber\\
&=&-gf^{abc}\, \frac{k^\sigma_2}{q^2}\, V_{\sigma}^{c}
\label{S}
\end{eqnarray}
and
and 
\begin{equation}
{\cal R}_{\mu}^{ab}\ =\ gf^{abc}\, V_{\mu}^{c}\, .
\label{R}
\end{equation}
Clearly, 
\begin{equation}
k_1^{\sigma}{\cal R}_{\sigma}^{ab}\ =\
-k_2^{\sigma}{\cal R}_{\sigma}^{ab}\ =\ q^2{\cal S}^{ab}.
\label{D2}
\end{equation}
We then have
\begin{eqnarray}
\Im m {\cal M}&=& \frac{1}{4}\, {\cal T}^{ab}_{\mu\nu}\, P^{\mu\sigma}
(k_1,\eta_1)\, P^{\nu\lambda}(k_2,\eta_2)\, {\cal T}^{ab*}_{\sigma\lambda}
\nonumber\\
&=& \frac{1}{4}\Big[ {{\cal T}_{s}}^{ab}_{\mu\nu}(\xi)+
{{\cal T}_{t}}^{ab}_{\mu\nu}\Big]\, P^{\mu\sigma}(k_1,\eta_1)\times\nonumber\\
&& P^{\nu\lambda}(k_2,\eta_2)\, \Big[ {{\cal T}_{s}}^{ab*}_{\sigma\lambda}(\xi)
+{{\cal T}_{t}}^{ab*}_{\sigma\lambda}\Big],
\label{MM}
\end{eqnarray}
where the polarization tensor $P^{\mu\nu}(k,\eta )$ is given by
\begin{equation}
P_{\mu\nu}(k,\eta )\ =\ -g_{\mu\nu}+ \frac{\eta_{\mu}k_{\nu}
+\eta_{\nu}k_{\mu} }{\eta k} + 
\eta^2 \frac{k_{\mu}k_{\nu}}{{(\eta k)}^2}\, .
\label{PhotPol}
\end{equation}
Moreover, we have that on-shell, {\em i.e.}, for $k^{2}=0$,
$k^{\mu}P_{\mu\nu}=0$. By virtue of  this last property, we see immediately
that if we write the three-gluon vertex of Eq.\ (\ref{3GV}) in the form 
\begin{eqnarray}
\Gamma_{\lambda\mu\nu} (q,-k_1,-k_2) &=& [(k_1-k_2)_{\lambda}g_{\mu\nu}+
2q_{\mu}g_{\lambda\nu}-2q_{\nu}g_{\lambda\mu}]\nonumber\\
&& +\ (-k_{1\mu}g_{\lambda\nu}+k_{2\nu}g_{\lambda\mu}) \nonumber\\
&=& \Gamma^F_{\lambda\mu\nu}(q,-k_1,-k_2)\, + \nonumber\\ 
&& \Gamma^P_{\lambda\mu\nu}(q,-k_1,-k_2)\, ,
\label{GFGP}
\end{eqnarray}
the term $\Gamma^P_{\rho\mu\nu}$ dies after hitting the polarization vectors
$P_{\mu\sigma}(k_1,\eta_1)$ and $P_{\nu\lambda}(k_2,\eta_2)$.  Therefore, if
we denote by ${\cal T}_{s}^{F}(\xi)$ the part of ${\cal T}_{s}$ which
survives, Eq.\ (\ref{MM}) becomes 
\bea
\Im m {\cal M}\ &=& \ \frac{1}{4}\,
\big[ {\cal T}_{s}^{F}(\xi)+{\cal T}_{t}\big]^{ab}_{\mu\nu}\,
P^{\mu\sigma}(k_1,\eta_1 )\times\nonumber\\ 
&& P^{\nu\lambda}(k_2,\eta_2  )\,
\big[ {\cal T}_{s}^{F}(\xi)
+{\cal T}_{t} \big]^{ab*}_{\sigma\lambda}\, .
\label{MM22}
\eea
The next step is to verify that any dependence on the GFP inside the
propagator $\Delta^{(\xi)}_{0\mu\nu}(q)$ of the off-shell gluon will
disappear. This is indeed so, because the longitudinal parts of
$\Delta_{0\mu\nu}$ either vanish because the external quark current is
conserved, or because they trigger the following EWI: 
\begin{equation}
q^{\mu}\Gamma^{F}_{\mu\alpha\beta}(q, -k_1, -k_2)\ =\ 
(k_1^2\ -\ k_2^2)g_{\alpha\beta}\, ,
\label{FWI}
\end{equation}
which vanishes on shell.  This last EWI is crucial, because in general,
current conservation alone is not sufficient to guarantee the GFP independence
of the final answer. In the covariant gauges for example, the gauge fixing term
is proportional to $q^{\mu}q^{\nu}$; current conservation kills such a term. 
But if we had chosen an axial gauge instead, {\em i.e.} 
\begin{equation}
\label{Deleta}
\Delta^{(\tilde{\eta})}_{0\mu\nu}(q)\ =\ \frac{P_{\mu\nu}
(q,\tilde{\eta})}{q^2}\, , 
\end{equation}
where $\tilde{\eta} \neq \eta$ in general, then only the term
${\tilde{\eta}_{\nu}}q_{\mu}$ vanishes because of current conservation,
whereas the term ${\tilde{\eta}_{\nu}}q_{\mu}$ can only disappear if Eq.\
(\ref{FWI}) holds.  So, Eq.\ (\ref{MM22}) becomes
\bea
\Im m {\cal M}\ &=&\ \frac{1}{4}
({\cal T}_{s}^{F}+{\cal T}_{t})^{ab}_{\mu\nu}\,
P^{\mu\sigma}(k_1,\eta_1)\times\nonumber\\ 
&& P^{\nu\lambda}(k_2,\eta_2)\,
({\cal T}_{s}^{F}+{\cal T}_{t})^{ab*}_{\sigma\lambda}\, ,
\label{MM2}
\eea
where the GFP-{\em independent} quantity ${\cal T}_{s}^{F}$ is given by
\begin{equation}
{{\cal T}_{s}}^{F,ab}_{\mu\nu}\ =\
-gf^{abc}\, \frac{g^{\rho\lambda}}{q^2}\, \,
\Gamma^{F}_{\lambda\mu\nu}(q,-k_1,-k_2)\, V_{\rho}^{c}\, .
\label{TsF}
\end{equation}
Next, we want to show that the dependence on $\eta_{\mu}$ and $\eta^2$
stemming from the polarization vectors disappears.  Using the on shell
conditions $k_1^2=k_2^2=0$, we can easily verify the following EWIs: 
\begin{eqnarray}
k_1^{\mu}{{\cal T}_{s}}^{F,ab}_{\mu\nu} & = &  
2 k_{2\nu}{\cal S}^{ab}\, -\, {\cal R}_{\nu}^{ab} \, ,
\nonumber\\
k_2^\nu {{\cal T}_{s}}^{F,ab}_{\mu\nu}  & = &  2 k_{1\mu}{\cal S}^{ab} 
\, +\, {\cal R}_{\mu}^{ab} \, , \nonumber\\
k_1^{\mu}{{\cal T}_{t}}^{ab}_{\mu\nu} & = & {\cal R}_{\nu}^{ab} \, ,
\nonumber\\
k_2^{\nu}{{\cal T}_{t}}^{ab}_{\mu\nu} & = & -{\cal R}_{\mu}^{ab} 
\label{w4}\, ,
\end{eqnarray}
from which we have that 
\begin{eqnarray}
k_1^{\mu}k_2^{\nu}{{\cal T}_{s}}^{F,ab}_{\mu\nu}& = & q^2{\cal S}^{ab}
\label{w5}\, ,\nonumber \\
k_1^{\mu}k_2^{\nu}{{\cal T}_{t}}^{ab}_{\mu\nu} &=& -q^2{\cal S}^{ab}\, .
\label{w6}
\end{eqnarray}

Using the above EWIs, it is now easy to check that indeed, all dependence on
both $\eta_{\mu}$ and $\eta^2$ cancels in Eq.\ (\ref{MM2}), as it should, and
we are finally left with (omitting the fully contracted colour and Lorentz
indices): 
\begin{eqnarray}
\Im m {\cal M} &=& \frac{1}{4}\, \Big[
\Big( {\cal T}_{s}^{F}{{\cal T}_{s}^{F}}^{*} -8 {\cal S}{\cal S}^{*}\Big) 
\nonumber\\
&&+ \Big( {\cal T}_{s}^{F}{\cal T}_{t}^{*} + {{\cal T}_{s}^{F}}^{*}
{\cal T}_{t} \Big) + {\cal T}_{t}{\cal T}_{t}^{*} \Big]\nonumber\\
&=& \Im m \widehat{{\cal M}}_1+ \Im m \widehat{{\cal M}}_2+
\Im m \widehat{{\cal M}}_{3} \, .\nonumber\\
&&{}
\label{MM3}
\end{eqnarray}
The first part is the genuine propagator-like piece, the second is the vertex,
and the third the box.  Employing the fact that
\begin{equation}
\Gamma^{F}_{\rho\mu\nu}\Gamma^{F,\mu\nu}_{\lambda} \ =\ 
-8q^2t_{\rho\lambda}(q) +
4{(k_1-k_2)}_{\rho}{(k_1-k_2)}_{\lambda}
\label{FRE}
\end{equation}
and 
\begin{eqnarray}
{\cal S}{\cal S }^{*} &=& g^2\, c_A\, V^c_\rho \, 
\frac{k_1^{\rho}k_1^{\lambda}}{(q^2)^2}\, V^{c}_{\lambda} \nonumber\\
&=& \frac{g^2}{4}\, c_A\, V^c_{\rho}\, \frac{(k_1-k_2)^\rho 
(k_1-k_2)^\lambda }{(q^2)^2}\, V^{c}_{\lambda}~,\nonumber\\
&&{}  
\end{eqnarray}
where $c_{A}$ is the eigenvalue of the Casimir operator in the adjoint
representation ($ c_A=N$ for SU$(N)$), we obtain for $\Im m \widehat{{\cal
M}}_1$
\bea
\Im m \widehat{{\cal M}}_1\ &=&\ \frac{g^2}{2}\, c_A
V^{c}_{\mu}\, \frac{1}{q^2}\, \Big[ -4q^2t^{\mu\nu}(q)\,\nonumber\\ 
&& +\,
{(k_1-k_2)}^{\mu}{(k_1-k_2)}^{\nu}\Big]\, \frac{1}{q^2}\, V^{c}_{\nu}\, .
\nonumber\\
&&{}
\eea 
This last expression must be integrated over the available 
two-body phase space; using standard results   
we arrive at the final
expression
\begin{equation}
\Im m \widehat{{\cal M}}_1\ =\
V^c_\mu\, \frac{1}{q^2}\, \Im m \widehat{\Pi}^{\mu\nu}(q)\,
\frac{1}{q^2}\, V^c_\nu\, ,
\end{equation}
with  
\begin{equation}
\label{IMQCD}
\Im m \widehat{\Pi}_{\mu\nu}(q)\ =\ -\, \frac{\alpha_s}{4}\, 
\frac{11c_A}{3}\, q^2 t_{\mu\nu}(q)\, ,
\end{equation}
and $\alpha_s=g^2/(4\pi)$. 
The vacuum polarization of the gluon within the PT is given by \cite{PT}
\begin{equation}
\label{PTgg}
\widehat{\Pi}_{\mu\nu}(q)\ =\ \frac{\alpha_s}{4\pi}\, \frac{11c_A}{3}\, 
t_{\mu\nu}(q)\, q^2\, \Big[\, \ln\Big(-\frac{q^2}{\mu^2}\Big)\, +\, C_{UV}\, 
\Big]\, .
\end{equation}
Here, $C_{UV}=1/\epsilon -\gamma_E + \ln 4\pi + C$, with $C$ being some
constant and $\mu$ is a subtraction point. In Eq.\ (\ref{PTgg}), it is
interesting to notice that a change of $\mu^2\to \mu '^2$ gives rise to a
variation of the constant $C$ by an amount $C'-C=\ln \mu'^2/\mu^2$.  Thus, a
general $\mu$-scheme renormalization yields
\begin{eqnarray}
\label{RPTgg}
\widehat{\Pi}_T^R (s) &=& \widehat{\Pi}_T (s)\, -\, (s-\mu^2)\Re e
\widehat{\Pi}_T'(\mu^2)\, -\, \Re e\widehat{\Pi}_T(\mu^2 ) \nonumber\\
&=& \frac{\alpha_s}{4\pi}\, \frac{11c_A}{3}\, s\, \Big[
\ln\Big(-\frac{s}{\mu^2}\Big)\, -\, 1\, +\, \frac{\mu^2}{s}\, \Big]\, .
\nonumber\\
&&
\end{eqnarray}
One can readily see now that $\Re e
\widehat{\Pi}^R_T(s)$ can be calculated by the following double 
subtracted dispersion relation:
\begin{equation}
\label{QCDDR}
\Re e \widehat{\Pi}^R_T(s)\ =\ \frac{(s-\mu^2)^2}{\pi}
\int\limits_0^\infty\, ds' \frac{\Im m \widehat{\Pi}_T(s')}{(s'-\mu^2)^2 
(s'-s)}\, .
\end{equation}
Inserting Eq.\ (\ref{IMQCD}) into Eq.\ (\ref{QCDDR}), it is not difficult to
show that it leads to the result given in Eq.\ (\ref{RPTgg}), a fact that 
demonstrates the analytic power of the dispersion relations.

\section{The electroweak case}

In this section, we will show how the same considerations apply directly to the
case of the electroweak sector of the SM. We consider the charged current
process $e^- \nu \to e^- \nu$ and assume that the electron mass $m_e$ is
non-zero, so that the external current is not conserved.  We focus on the part
of the amplitude which has a threshold at $q^2=M^2_{W}$. This corresponds the
virtual process $W^-\to W^-\gamma$, where $\gamma$ is the photon. From the
optical theorem, we have 
\begin{equation}
\Im m \langle e^{-}\nu|T|e^{-}\nu\rangle\ =\ \frac{1}{2}
\int dX_{LIPS}\ |\langle e^{-}\nu|T|W^{-}\gamma\rangle|^{2} \, .
\label{OTWg}
\end{equation}

We set again ${\cal M}=\langle e^{-}\nu|T|e^{-}\nu\rangle$ and ${\cal T}=
\langle e^{-}\nu|T|W^{-}\gamma\rangle$. As in the case of QCD,  the amplitude
consists of two distinct parts, a part that contains an electron propagator
(Fig. 4(a)) and a part that does not, which is shown in Figs. 4(b) and 4(c). 
As before, we denote them by ${\cal T}_t$ and ${\cal T}_s(\xi_w )$, 
respectively. We first define 
\begin{equation}
V_L^\mu\ =\ \frac{g_w}{2\sqrt{2}}\,
\bar{v}(p_2)\gamma^{\mu}(1-\gamma_{5})u(p_1) 
\end{equation}
and 
\begin{equation}
S_R\ =\ \frac{g_w}{2\sqrt{2}}\, \frac{m_e}{M_{W}}\,
\bar{v}(p_2)(1+\gamma_{5})u(p_1)\, .
\end{equation}
Clearly, one has the EWI
\begin{equation}
q_{\mu}V_L^{\mu}\ =\ M_W S_R\, .
\label{VVWI}
\label{VV}
\end{equation}
The amplitude ${\cal T}_{s}$ can the be written down in the closed form
\bea
\label{Tsxiw}
{{\cal T}_{s}}_{\mu\nu}(\xi_{w})\ &=&\ i
V_L^\lambda\, \Delta_{0\lambda}^{(\xi_{w}),\rho}(q)\,
\Gamma^{\gamma W^-W^+}_{\nu\rho\mu}\ + \nonumber\\
&& \ iS_R\, 
D_0^{(\xi_{w})}(q)\, \Gamma^{\gamma G^- W^+} _{\nu\mu}\, ,
\eea
where $\Gamma^{\gamma W^-W^+}_{\nu\rho\mu}=e \Gamma_{\nu\rho\mu}
(-k_2,q,-k_1)$ is the tree-level $\gamma W^- W^+$ vertex and $\Gamma^{\gamma
G^- W^+}_{\nu\mu} = eM_Wg_{\mu\nu}$ is the tree-level $\gamma G^-W^+$ vertex.
In the expression (\ref{Tsxiw}), we explicitly display the dependence on the
GFP $\xi_{w}$. In addition, the amplitude ${\cal T}_{t}$ is given by 
\begin{equation}
{\cal T}_{t}^{\mu\nu}\ =\ \frac{ieg_w}{2\sqrt{2}}\,
\bar{v}(p_2)\, \gamma^\mu (1-\gamma_{5})\, 
\frac{1}{\not\! p_1-\not\! k_2 -m_e}\,  \gamma^\nu\,  u(p_1)\, . 
\end{equation}
Notice that ${\cal T}_{t}^{\mu\nu}$ does not depend on $\xi_{w}$. Denoting by
$k_1$ the four-momentum of the $W$ and by $k_2$ that of the photon, Eq.\
(\ref{OTWg}) becomes
\begin{equation}
\label{Mew}
\Im m {\cal M}\ =\  
{\cal T}_{\mu\nu}Q^{\mu\rho}(k_1)P^{\nu\sigma}(k_2,\eta)
{\cal T}^*_{\rho\sigma}\, ,
\end{equation}
where $P^{\mu\nu}$ is the photon polarization tensor given in Eq.\ 
(\ref{PhotPol}), and
\begin{equation}
\label{WPol}
Q^{\mu\nu} (k)\ =\ -g^{\mu\nu}\, +\, \frac{k^\mu k^\nu}{M^2_W}\, 
\end{equation}
is the $W$ polarization tensor. The polarization tensor $Q^{\mu\nu} (k)$
shares the property that, on shell, {\em i.e.},  for $k^2= M^2_{W}$, $k^{\mu}
Q_{\mu\nu}(k)=0$. Furthermore, in Eq.\ (\ref{Mew}), we omit the integration
measure $1/2\int dX_{LIPS}$. 
\begin{equation}
\label{Prop1}
\Delta_{0\mu\nu}^{(\xi_Q)}(q)\ =\ t_{\mu\nu}(q)\, \frac{1}{q^2-M^2}\, -\, 
\ell_{\mu\nu}(q)\, \frac{\xi_Q}{q^2-\xi_Q M^2}\ ,
\end{equation}
with
\begin{displaymath} 
t_{\mu\nu}(q)\ =\ {}- g_{\mu\nu} + \frac{q_\mu q_\nu}{q^2}\, ,\quad
\ell_{\mu\nu}(q)\ =\ \frac{q_\mu q_\nu}{q^2}\ .
\end{displaymath}
First, we will show how the dependence on the GFP $\xi_{w}$ cancels. To that
end, we employ the usual decomposition 
\begin{equation}
\Delta_{0\mu\nu}^{(\xi_{w})}(q) \
=\  U_{\mu\nu}(q)\ -\ \frac{q_\mu q_\nu}{M_W^2} D_0^{(\xi_{w})}(q^2)\, ,
\label{D0xi}
\end{equation}
the EWI  
\bea
&& q^{\rho}\Gamma^{\gamma W^- W^+}_{\nu\rho\mu}(-k_2,q,-k_1)\, 
Q^{\mu\lambda}(k_1)P^{\nu\sigma}(k_2,\eta)\ =\ 
\nonumber\\
&& M_W \Gamma^{\gamma G^- W^+}_{\mu\nu}\, Q^{\mu\lambda}(k_1)
P^{\nu\sigma}(k_2,\eta)
\eea
and the EWI of Eq.\ (\ref{VVWI}), and we obtain the following 
$\xi_{w}$-independent expression for ${\cal T}_{s}^{\mu\nu}$ 
\bea
\label{Tsinter}
{\cal T}_{s}^{\mu\nu}&=&
ieV_L^\lambda U_{\lambda\rho}(q)\Gamma^{\nu\rho\mu}(-k_2,q,-k_1)
\nonumber\\
\ &=&\ 
ieV_L^{\lambda}\, U_{\lambda\rho}(q)\, 
\Gamma^{F,\nu\rho\mu}(-k_2,q,-k_1)\nonumber\\
&=& {{\cal T}_{s}}^{F,\mu\nu}\, ,
\eea
where contraction over the polarization tensors $Q_{\mu\nu}$ and $P_{\mu\nu}$
is implied. In the last step of Eq.\ (\ref{Tsinter}), we have used the fact
that the $\Gamma^{P}$ part of the vertex, defined in Eq.\ (\ref{GFGP}),
vanishes when contracted with the polarization tensors. 

Next, we show how the dependence on the four-vector $\eta_{\mu}$ and the
parameter $\eta^2$ vanishes. First, it is straightforward to verify the
following EWI:
\begin{eqnarray}
k^\mu_1 \Gamma_{\nu\rho\mu}^F & = & 
[U^{-1}_\gamma (k_2) - U^{-1}(q)- U^{-1}(k_1)]_{\nu\rho}\nonumber\\
&&+2M^2_W g_{\nu\rho} + (k_1 - k_2 )_\nu {k_1}_\rho \nonumber\\ 
&=&-U_{\nu\rho}^{-1}(q)\, +\, 2M^2_W g_{\nu\rho}\, -\,  \nonumber\\
&& k_{2\nu} (k_1 - k_2 )_\rho \, ,
\label{WIx}
\end{eqnarray}
where the on-shell conditions $k_1^2=M^2_W$ and $k_2^2=0$ are used in the last
equality of Eq.\ (\ref{WIx}). Similarly, one has
\begin{eqnarray}
k^\nu_2 \Gamma_{\nu\rho\mu}^F& = & 
[U^{-1}(q)- U^{-1}(k_1) +U^{-1}_\gamma (k_2)]_{\rho\mu}\nonumber\\
&&+ k_{2\rho} (k_1 - k_2 )_\mu \nonumber\\
&=&U_{\rho\mu}^{-1}(q)\, - \, (k_1 - k_2 )_\rho k_{1\mu}\, ,
\label{WIy} 
\end{eqnarray}
with
\begin{eqnarray}
U_{\alpha\beta}^{-1}(q)& =& (q^2-M^2_W)\, t_{\alpha\beta}\ +\
M^2_W\, \ell_{\alpha\beta}\, ,\nonumber\\
{U^{-1}_\gamma}_{\alpha\beta}(q)&=& q^2\, t_{\alpha\beta}\, .
\end{eqnarray}
So, when the $\eta^{\sigma}k_2^{\nu}$ term from $P_{\nu\sigma}(k_2,\eta )$
gets contracted with ${\cal T}_{\mu\nu}$, we have 
\begin{eqnarray}
\eta^\sigma k_2^\nu {{\cal T}_{s}}_{\mu\nu} &=& 
i e\eta^\sigma V_L^{\lambda}\Big[ g_{\lambda\mu}\ -\ 
U^\alpha_\lambda(q)\, U^{-1}_{\alpha\mu} (k_1)\Big], \nonumber\\
\eta^{\sigma}k_2^{\nu}{{\cal T}_{t}}_{\mu\nu} & =& 
-i e\eta^{\sigma}V_{L\mu}\, .
\end{eqnarray}
Adding the last two equations by parts, we find
\begin{equation} 
\eta^\sigma k_2^\nu {\cal T}_{\mu\nu}\ =\  i e\eta^\sigma
V_{L}^\lambda\, U^\alpha_\lambda(q)\, U^{-1}_{\alpha\mu} (k_1)\, .
\end{equation}
Since the result is proportional to ${k_1}_{\mu}$, the four-momentum of the
external $W$ boson, we immediately see that 
\begin{equation}
\eta^{\sigma}k_2^{\nu}{\cal T}_{\mu\nu}Q^{\mu\rho}(k_1)\ =\ 0\, .
\end{equation}
For the same reasons, the term proportional to $\eta^2$ vanishes as well. 
Consequently, $\Im m {\cal M}$ takes on the form
\begin{eqnarray}
\Im m {\cal M}&=& -({\cal T}_{s}^{F}+{\cal T}_{t})_{\mu\nu}Q^{\mu\rho}(k_1)
( {\cal T}_{s}^{F}+{\cal T}_{t})^{*}_{\rho\nu}\nonumber\\
&=& ( {\cal T}_{s}^{F}+{\cal T}_{t})^{\mu\nu}
( {\cal T}_{s}^{F}+{\cal T}_{t})^{*}_{\mu\nu}\ - \nonumber\\
&& ( {\cal T}_{s}^{F}+{\cal T}_{t})^{\mu\nu}
\, \frac{k_{1\mu} k_1^\rho }{M_W^2 }\, 
( {\cal T}_{s}^{F}+{\cal T}_{t})^{*}_{\rho\nu}\nonumber\\
&=&\Im m {\cal M}^{a}+\Im m {\cal M}^{b} .
\label{XYZ}
\end{eqnarray}
The absorptive sub-amplitude, $\Im m {\cal M}^{a}$, consists of three terms, 
\begin{eqnarray}
\Im m {\cal M}^{a} &=&{\cal T}_{s}^{F}{{\cal T}_{s}^{F}}^{*}+
( {\cal T}_{s}^{F}{\cal T}_{t}^{*}+  {\cal T}_{t} {{\cal T}_{s}^{F}}^{*})
+ {\cal T}_{t}{\cal T}_{t}^{*}\nonumber\\
&=& \Im m \widehat{{\cal M}}^{a}_1+
\Im m \widehat{{\cal M}}^{a}_2+\Im m \widehat{{\cal M}}^{a}_{3}\, .
\end{eqnarray}
The first term, $\Im m \widehat{{\cal M}}^{a}_1$, can easily be identified 
with a propagator-like contribution. In particular, using Eq.\ (\ref{FRE}), 
we find 
\bea
\Im m \widehat{{\cal M}}^{a}_1\ &=&\ e^2\, 
V_L^{\rho}\, U_{\rho\mu}(q)\,
\Big[ -8 q^2t^{\mu\nu}(q)\, +\,\times \nonumber\\
&& 4{(k_1-k_2)}^{\mu}{(k_1-k_2)}^{\nu} \Big]\times\nonumber\\
&& U_{\nu\lambda}(q)V_L^{\lambda}\, .
\label{IMA1}
\eea
The amplitudes, $\Im m \widehat{{\cal M}}^{a}_2$ and 
$\Im m \widehat{{\cal M}}^{a}_{3}$, are vertex- and box-like contributions, 
respectively, and they will not be considered any further here. 

We must now isolate the corresponding propagator-like piece from $\Im m {\cal
M}^{b}$. By virtue of the EWI of Eq.\ (\ref{WIx}), we have
\bea
k_1^{\mu}{{\cal T}_{s}}^{F}_{\mu\nu}\ &=&\ -ieV_{L\nu}\, -\, ie 
V_{L\lambda}\, U^{\lambda\rho}(q)\,\times \nonumber\\ 
&& \Big[ (k_1-k_2)_\rho {k_2}_\nu
- 2M^2_Wg_{\rho\nu} \Big]\, .
\label{A1}
\eea
In addition, we evaluate the EWI
\begin{eqnarray}
k_1^{\mu}{{\cal T}_{t}}_{\mu\nu} &=& 
ieV_{L\nu}\, +\, M_W\, \frac{ieg_wm_e}{2\sqrt{2}M_W}\, 
\bar{v}(p_2)\, (1+\gamma_5)\, \nonumber\\ 
&&\frac{1}{\not\! p_1 -\not\! k_2-m_e}\, 
\gamma_\nu\,  u(p_1)
\nonumber\\ 
&=& ieV_{L\nu}\ + \nonumber\\
&& \ M_W{\cal L}_{\nu}\, ,
\label{Gbox}
\end{eqnarray}
which is shown diagrammatically in Fig.\ 5. 

Adding Eqs.\ (\ref{A1}) and (\ref{Gbox}) by parts, we obtain
\bea
k_1^{\mu}\, ( {\cal T}_{s}^{F}+{\cal T}_{t})_{\mu\nu}\ &=&\
-ieV_{L\lambda}\, U^{\lambda\rho}(q)\times\nonumber\\
&& \Big[(k_1-k_2)_{\rho}{k_2}_{\nu} - 2M^2_{W}g_{\rho\nu}
 \Big]\ +\ M_W{\cal L}_{\nu}\, .\nonumber\\
&&{}
\eea
Making now use of the EWI of Eq.\ (\ref{VV}) and writing 
\begin{equation}
S_R\ =\ M_W V_{L\mu}\, U^{\mu\nu}(q)\, q_\nu 
\end{equation} 
yields the following WI for ${\cal L}_{\sigma}$:
\begin{equation}
k_2^\nu\, {\cal L}_\nu\ =\ -ie S_R \ =\
-ieM_W\, V_{L\alpha}\, U^{\alpha\beta}(q)\, q_{\beta}\, .
\end{equation}

Taking the above relations into account, we eventually obtain
\begin{eqnarray}
\Im m {\cal M}^{b}&=& -e^2\, V_{L\rho}\, U^{\rho\mu}(q)\, 
\Big[ 4M^2_W g_{\mu\nu}+ \nonumber\\
&& 2 (k_1-k_2)_\mu(k_1-k_2)_\nu \Big] 
U^{\nu\lambda}(q)\, V_{L\lambda}\nonumber\\
&& -2ieM_W\, \Big[ V_{L\rho}\, U^{\rho\nu}(q)\, {\cal L}_{\nu}^{*} \,
-\, {\cal L}_{\nu}\, U^{\nu\lambda}(q)\, V_{L\lambda} \Big]
\nonumber\\
&& -{\cal L}^{\nu}{\cal L}_{\nu}^{*}\nonumber\\
&=& \Im m \widehat{{\cal M}}^{b}_1 + 
\Im m \widehat{{\cal M}}^{b}_2 + 
\Im m \widehat{{\cal M}}^{b}_{3}\, .
\label{IMA2}
\end{eqnarray}
Adding the two propagator-like parts $\Im m \widehat{{\cal M}}^{a}_1$ and 
$\Im m \widehat{{\cal M}}^{b}_1$  from Eqs.\ (\ref{IMA1}) and (\ref{IMA2}), 
respectively, we find
\begin{eqnarray}
\Im m \widehat{{\cal M}}_1&=& 
\Im m \widehat{{\cal M}}^{a}_1+ \Im m \widehat{{\cal M}}^{b}_1\nonumber \\
&=& e^2\, V_L^\rho\, U_{\rho\mu}(q)
\Big[ -8q^2t^{\mu\nu}(q)\, -\, 4M_{W}^2g^{\mu\nu}\nonumber \\
&& +\, 2{(k_1-k_2)}^{\mu}{(k_1-k_2)}^{\nu}\Big] U_{\nu\lambda}(q)\, 
V_L^{\lambda}.
\nonumber\\
&&
\label{IMA}
\end{eqnarray}  
Next, we carry out the phase-space integration over $1/2\int dX_{LIPS}$, 
using standard integration formulae, 
we have
\begin{equation}
\Im m \widehat{{\cal M}}_1\ =\
V_{L\rho}U^{\rho\mu}(q)\ \Im m \widehat{\Pi}_{\mu\nu}^{W}\
U^{\nu\lambda}(q)\, V_{L\lambda}\, ,
\end{equation}
with
\begin{eqnarray}
\Im m \widehat{\Pi}_{\mu\nu}^{W}(q) &=& \Im m \widehat{\Pi}_T^{W}(q^2)\,
t_{\mu\nu}(q)\ + \nonumber\\
&& \ \Im m \widehat{\Pi}_L^{W}(q^2)\, 
\ell_{\mu\nu}(q),\nonumber\\
\Im m \widehat{\Pi}_T^{W}(q^2) &=& \frac{\alpha_{em}}{2}\, (q^2-M^2_W)\times
\nonumber\\
&&\Big( -\frac{11}{3}\, +\, \frac{4M^2_W}{3q^2}\, +\, \frac{M^4_W}{3q^4}
\Big)\, ,\nonumber\\
\Im m \widehat{\Pi}_L^{W}(q^2) &=& \frac{\alpha_{em}}{2}\, (q^2-M^2_W)\times
\nonumber\\
&& \Big( -\, \frac{2M^2_W}{q^2}\, +\, \frac{M^4_W}{q^4}
\Big)\, .
\end{eqnarray}
Here, $\alpha_{em}=e^2/(4\pi)$ is the electromagnetic fine structure constant.
The real part of the transverse, on-shell renormalized, $W$-boson self-energy,
$\Re e\widehat{\Pi}_T^{W,R}(s)$, can be determined by means of a doubly
subtracted dispersion relation. Furthermore, we have to assume a
fictitious photon mass, $\mu_\gamma$, in order to regulate the infra-red (IR) 
divergences. More explicitly, the relevant dispersion relation reads
\begin{eqnarray}
\label{RDRW}
\Re e\widehat{\Pi}_T^{W,R}(s) &=& \Re e\widehat{\Pi}_T^W (s)\, -\,
(s-M^2_W)\Re e\widehat{\Pi}_T^W{}'(M^2_W)\, \nonumber\\
&& -\, \Re e\widehat{\Pi}_T^W(M^2_W)
\nonumber\\
&=& 
\frac{(s-M^2_W)^2}{\pi} \nonumber\\
&& \int\limits^{\infty}_{(M_W+\mu_\gamma )^2}
\, \frac{ ds'\, \Im m \widehat{\Pi}^W_T(s')}{(s'-M^2_W)^2 (s'-s)}\, .
\end{eqnarray}
To obtain the analytic form of $\Re e\widehat{\Pi}_T^{W,R}(s)$, we first
evaluate the following integrals:
\begin{eqnarray}
\label{F0}
F_0(s) &=& (s-M^2_W) \int\limits^\infty_{(M_W+\mu_\gamma )^2}
\frac{ds'}{(s'-M^2_W) (s'-s)}\nonumber\\
&=&-\, \ln\Big(\, \frac{|s-M^2_W|}{2M_W\mu_\gamma}\, \Big)\, ,\\[0.3cm]
\label{F1}
F_1(s) &=& (s-M^2_W) \int\limits^\infty_{(M_W+\mu_\gamma )^2}
\frac{ds'}{(s'-M^2_W) (s'-s)}\, \frac{M^2_W}{s'}\nonumber\\
&=&-\, \frac{M^2_W}{s}\, \ln\Big(\, \frac{|s-M^2_W|}{2M_W\mu_\gamma}\, \Big)\,
-\,\nonumber\\ 
&&\Big(1-\frac{M^2_W}{s}\Big)\ln\Big(\frac{M_W}{2\mu_\gamma}\Big)\, ,
\\[0.3cm]
\label{F2}
F_2(s) &=& (s-M^2_W) \int\limits^\infty_{(M_W+\mu_\gamma )^2}
\frac{ds'}{(s'-M^2_W) (s'-s)}\, \frac{M^4_W}{s'^2} \nonumber\\
&=&-\, \frac{M^4_W}{s^2}\, \ln\Big(\, \frac{|s-M^2_W|}{2M_W\mu_\gamma}\, 
\Big)\, - \nonumber\\
&&\ln\Big(\frac{M_W}{2\mu_\gamma}\Big)\, +\, 1\, -\, 
\frac{M^2_W}{s}\, ,
\end{eqnarray}
Using the integrals defined in Eqs.\ (\ref{F0})--(\ref{F2}), one then
obtains
\begin{equation}
\label{RenWPT}
\Re e\widehat{\Pi}_T^{W}(s)\ =\ \frac{\alpha_{em}}{2}\, (s-M^2_W)\,
\Big( -\frac{11}{3}F_0\, +\, \frac{4}{3}F_1\, +\, \frac{1}{3}F_2 \Big)\, .
\end{equation}
Eq.\ (\ref{RenWPT}) coincides with the PT $W$-boson self-energy \cite{JPs}
or equivalently with the $W$-boson self-energy computed in the 
background field method \cite{BFM}
for 
$\xi_Q=1$ \cite{DDW}.

\section{The Higgs boson resonance}

When  the    center-of-mass  (c.m.)   energy   $\sqrt{s}$
approaches $M_H$,  amplitudes  containing  an  $s$-channel Higgs boson
become singular,  and must be  regulated.  The naive  extension of the
standard Breit-Wigner procedure  to    this  case would consist     of
replacing the  free  Higgs  boson  propagator $\Delta_H  (s) =    (s -
M^2_H)^{-1}$ by   a resummed  propagator  of  the form   $[s - M^2_H +
\Pi^{HH}(s)]^{-1}$, where  $\Pi^{HH}(s)$  is the one-loop  Higgs boson
self-energy.   However,  bosonic   radiative corrections  induce    an
additional   dependence on  the GFP,  as  one  can verify  by explicit
calculations   in   a variety   of conventional gauges,    such as the
renormalizable  ($R_{\xi}$),   or  axial  gauges.   
Turning   to  more
elaborate gauge fixing  schemes does not improve  the  situation.  For
example,     within  the Background  Field   Method 
the contribution of  the   $Z$
boson-loop reads: \cite{PPHiggs}
\begin{eqnarray}
  \label{DBFG}
\Pi^{\widehat{H}\widehat{H}}_{(ZZ)}(s,\xi_Q)\ &=&\ \frac{\alpha_w}{32\pi}\,
\frac{s^2}{M^2_W}\, \Big\{
\Big(\, 1\, -\, 4\, \frac{M^2_Z}{s}\, +\, \nonumber\\
&& 12\, \frac{M^4_Z}{s^2}\Big) B_0(s,M^2_Z,M^2_Z) - \nonumber\\
&& \Big[ 1\, +\, 4\xi_Q\, \frac{M^2_Z}{s}\, \nonumber\\
&& -\, (M^2_H+4\xi_QM^2_Z)\,
\frac{M^2_H}{s^2}\, \Big]\nonumber\\
&&\times B_0(s,\xi_Q M^2_Z,\xi_Q M^2_Z)\, \Big \}\, ,
\nonumber\\
&&{}
\end{eqnarray}
where $\alpha_w=g^2_w/(4\pi)$ is the weak  fine structure constant and
$B_0$ is  the usual Passarino-Veltman  function \cite{tHV}.   
The presence of the
GFP $\xi_Q$ results in bad high energy   behaviour and the appearance of
unphysical  thresholds,  as   can  be verified  directly    using $\Im
mB_0(s,M^2,M^2) =  \theta(s-4M^2) \pi (1-4M^2/s)^{1/2}$.   Even though
to  any  order   in perturbation    theory  physical  amplitudes   are
GFP-independent,  and   display   only physical  thresholds, resumming
$\Pi^{\widehat{H}  \widehat{H}}_{(ZZ)}   (s,\xi_Q)$     will introduce
artifacts   to the  resonant  amplitude.   Even in   the unitary gauge
($\xi_Q \to\infty$),   where only   physical  thresholds survive,  the
$s^2$-growth in Eq.\ (\ref{DBFG})  grossly contradicts the equivalence
theorem.

As explained above,
in  the PT framework a  modified one-loop self-energy for the
Higgs boson  can  be constructed,  by   appending to the  conventional
self-energy additional  propagator-like contributions concealed inside
vertices and      boxes.   These contributions      can  be identified
systematically, by resorting exclusively to elementary Ward identities
of  the  form $\not\!  k  (v  + a\gamma_5) = (\not\!     k + \not\!  p
-m)(v+a\gamma_5)-  (v-a\gamma_5)   (\not\!  p  -m)   + 2a  m\gamma_5$,
triggered by the longitudinal virtual momenta $k_\mu$.  Following this
procedure, we find the PT Higgs-boson self-energy \cite{PPHiggs}
\bea
  \label{HPT}
\widehat{\Pi}^{HH}_{(ZZ)}(s)\ &=&\ \frac{\alpha_w}{32\pi}\frac{M_H^4}{M_W^2}
\Big[\, 1+4\frac{M_Z^2}{M_H^2}-\nonumber\\
&& 4\frac{M_Z^2}{M_H^4}
(2s - 3M_Z^2)\, \Big]B_0(s,M^2_Z,M^2_Z)\, , \nonumber\\
&&{}
\eea
which  is   GFP-independent in any   gauge   fixing scheme,  universal
\cite{NJW0}, grows linearly with $s$,  and displays physical thresholds
only.  

To     verify    that
$\widehat{\Pi}^{HH}_{(ZZ)}(s)$  satisfies the optical theorem
 {\it  individually}
consider    the
tree-level  transition   amplitude ${\cal   T}(ZZ)$  for the   process
$f(p_1)\bar{f}(p_2)\to  Z(k_1)Z(k_2)$; it is the  sum of an $s$- and a
$t$-  channel contribution, denoted   by ${\cal T}^H_s(ZZ)$ and ${\cal
  T}_t(ZZ)$, respectively, given by
\begin{eqnarray}
  \label{THsZZ}
{\cal T}^H_{s\, \mu\nu} (ZZ)\ &=&\ \Gamma^{HZZ}_{0\mu\nu}\, 
\Delta_H (s)\ \bar{v}(p_2) \Gamma^{Hf\bar{f}}_0 u(p_1)\, ,\nonumber\\
{\cal T}_{t\, \mu\nu} (ZZ)\ &=&\ \bar{v}(p_2)\Big( 
\Gamma^{Zf\bar{f}}_{0\nu}\, \frac{1}{\not\! p_1 + \not\! k_1 - m_f}\, 
\Gamma^{Zf\bar{f}}_{0\mu}\, +\,\nonumber\\
&& 
\Gamma^{Zf\bar{f}}_{0\mu}\, \frac{1}{\not\! p_1 + \not\! k_2 - m_f}\,
                            \Gamma^{Zf\bar{f}}_{0\nu} \Big)u(p_1)\, .
\nonumber\\
&&{}
\end{eqnarray}
Here,   $s=(p_1+p_2)^{2}=(k_1+k_2)^2$  is the   c.m.\  energy squared,
$\Gamma^{HZZ}_{0\mu\nu}   =  ig_w\,     M^2_Z  /  M_W     g_{\mu\nu}$,
$\Gamma^{Hf\bar{f}}_0   =    -i    g_w\,  m_f    /   (2    M_W)$   and
$\Gamma^{Zf\bar{f}}_{0\mu}=g_w/(2ic_w)\, \gamma_\mu\, [  T^f_z
(1 - \gamma_5) - 2Q_fs^2_w]$, with $c_w = \sqrt{1 - s^2_w} = M_W/M_Z$,
are  the   tree-level  $HZZ$,  $Hf\bar{f}$  and $Zf\bar{f}$ couplings,
respectively, and $Q_f$ is the electric charge of the fermion $f$, and
$T^f_z$ its $z$-component of the weak isospin.   We then calculate the
expression  $[{\cal  T}^H_{s\, \mu\nu}  (ZZ) \\
+  {\cal  T}_{t\, \mu\nu}
(ZZ)]  Q^{\mu\rho}(k_1)Q^{\nu\sigma}(k_2) [{\cal T}^H_{s\, \rho\sigma}
(ZZ)+{\cal     T}_{t\,   \rho\sigma} (ZZ)]^*$,  where  $Q^{\mu\nu}(k)=
-g^{\mu\nu}+k^{\mu}k^{\nu}/M^2_Z$    denotes  the usual   polarization
tensor, and   isolate  its Higgs-boson mediated  part.   To accomplish
this,  one must first    use  the longitudinal  momenta   coming  from
$Q^{\mu\rho}(k_1)$  and  $Q^{\nu\sigma}(k_2)$ in  order to extract the
Higgs-boson part of ${\cal T}_{t}^{\mu\nu}(ZZ)$, {\em i.e.},
\bea
 \label{TP}
\frac{k_1^\mu k_2^\nu }{M^2_Z}\, {\cal T}_{t\,\mu\nu}(ZZ)\ &=&\
{\cal T}_{P}^{H}+ \dots\ \nonumber\\
{\cal T}_{P}^{H} &=&\ -\, \frac{ig_w}{2M_W}\ 
\bar{v}(p_2)\Gamma^{Hf\bar{f}}_0 u(p_1)\  \, ,\nonumber\\
&&{}
\eea
where   the   ellipses denote   genuine  $t$-channel  (not Higgs-boson
related)  contributions.  Then,  one    must append the   piece ${\cal
  T}_{P}^{H}$  ${\cal  T}_{P}^{H*}$ to   the ``naive'' Higgs-dependent
part    ${\cal    T}^H_{s\,    \mu\nu}   (ZZ)$      $Q^{\mu\rho}(k_1)$
$Q^{\nu\sigma}(k_2)$      ${\cal   T}^{H*}_{s\,  \rho\sigma}    (ZZ)$.
Integrating the expression so  obtained over the two-body phase space,
we finally arrive at the imaginary  part of Eq.\ (\ref{HPT}), which is
the announced result.

The   gauge-invariance of  the   $S$ matrix imposes  tree-level   Ward
identities  on  the unrenormalized    one-loop  PT Green's   functions
\cite{PT,PT2}.  The requirement that the same Ward identities should be
maintained {\it after} 
renormalization gives rise to important QED-type relations
for the  renormalization constants  of the  theory.   Specifically, we
find
\bea
  \label{ZWZH}
&& \widehat{Z}_W\ =\ \widehat{Z}_{g_w}^{-2}\, ,\quad
\widehat{Z}_Z\ =\  \widehat{Z}_W \widehat{Z}_{c_w}^2\, ,\nonumber\\
&& \widehat{Z}_H\ =\ \widehat{Z}_W\, (1+ \delta M^2_W/M^2_W)\, ,
\eea 
where $\widehat{Z}_W$,   $\widehat{Z}_Z$, and $\widehat{Z}_H$  are the
wave-function renormalizations of  the   $W$,  $Z$ and  $H$    fields,
respectively, $\widehat{Z}_{g_w}$ is the coupling renormalization, and
$\widehat{Z}_{c_w}=(1+\delta    M_W^2/M_W^2)^{1/2}(1+\delta  M_Z^2   /
M_Z^2)^{-1/2}$. The  renormalization of the bare  resummed Higgs-boson
propagator $\hat{\Delta}^{H,0}(s)$ proceeds as follows:
\bea
  \label{DeltH}
\hat{\Delta}^{H,0}(s)\, &=&\, [\,s - (M^0_H)^2 +
\widehat{\Pi}^{HH,0}(s) ]^{-1}\, \nonumber\\
&=&\, \widehat{Z}_H [\,s - M_H^2 +
\widehat{\Pi}^{HH}(s)]^{-1}\, \nonumber\\
&=&\, \widehat{Z}_H\,
\hat{\Delta}^H(s)\, , 
\eea
with $(M^0_H)^2 = M_H^2 +  \delta M_H^2$. The renormalized Higgs-boson
mass $M_H^2$   may be defined  as the  real part  of  the complex pole
position    of  $\hat{\Delta}^H(s)$.   
Employing the
relations in Eq.\ (\ref{ZWZH}), we observe that the universal quantity
\bea
  \label{RGIC}
  \widehat{R}^{H,0}(s)\ & = &\  \frac{(g^0_w)^2}{(M^0_W)^2}\,
  \hat{\Delta}^{H,0}(s) \ =\  \frac{g^2_w}{M_W^2}\, \hat{\Delta}^{H}
  (s)\nonumber\\ 
&=&\ \widehat{R}^H(s)
\eea
is   invariant  under  the  renormalization   group.   This  important
universal  property of the Higgs boson  is  true for non-Abelian gauge
theories with spontaneous symmetry  breaking (SSB), but does not  hold
in general.

An  additional,  highly non-trivial  constraint,  must  be  imposed on
resummed amplitudes; they have  to obey the (generalized)  equivalence
theorem (GET),  which  is known to  be  satisfied  before resummation,
order by order in perturbation  theory.   For the specific example  of
the amplitude ${\cal T} (ZZ)={\cal T}^H_s+{\cal T}_t$ , the GET states
that
\bea
  \label{GETZZ}
{\cal T}(Z_LZ_L)\ &=&\  -\, {\cal T}(G^0G^0) \,  
-i\,{\cal T}(G^0 z)\,\nonumber\\ 
&&-i \,{\cal T}(z G^0)\, +\, {\cal T}(z z)\, ,
\eea
where $Z_L$ is  the longitudinal component  of the $Z$ boson, $G^0$ is
its  associated    would-be    Goldstone boson,  and     $z^\mu  (k) =
\varepsilon^\mu_L(k) - k^\mu/M_W$ is the energetically suppressed part
of  the longitudinal polarization  vector  $\varepsilon^\mu_L$.  It is
crucial  to observe,  however, that already    at the tree level,  the
conventional $s$- and  $t$- channel sub-amplitudes  ${\cal T}^H_s$ and
${\cal T}_t$ fail  to satisfy the GET  individually.  To  verify that,
one   has  to   calculate ${\cal   T}_s^H  (Z_LZ_L)$,  using  explicit
expressions for the   longitudinal polarization vectors, and check  if
the  answer obtained is  equal to the Higgs-boson mediated $s$-channel
part  of the LHS  of Eq.\ (\ref{GETZZ}).   In particular, in the c.m.\
system, we have $z^\mu (k_1) = \varepsilon^\mu_L (k_1) - k_1^\mu/M_Z =
- 2M_Zk^\mu_2/s    +  {\cal  O}(M^4_Z/s^2)$,  and exactly    analogous
expressions  for $z^\mu (k_2)$.  The  residual  vector $z^\mu (k)$ has
the properties    $z_\mu k^\mu  =  -M_Z$  and  $z^2  = 0$.     After a
straightforward calculation, we obtain ${\cal T}_s^H (Z_LZ_L) = -{\cal
  T}^H_s (G^0G^0) -i {\cal T}^H_s (zG^0) -i {\cal T}^H_s(G^0z) + {\cal
  T}^H_s (zz) - {\cal T}^H_P$, where
\be
{\cal T}^H_s (G^0G^0)\ =\  \Gamma_0^{HG^0G^0}\,  
\Delta_H (s)\ \bar{v}(p_2)\Gamma^{Hf\bar{f}}_0 u(p_1)\, ,
\nonumber
\ee
\bea
{\cal T}^H_s (zG^0)\, +\, {\cal T}^H_s(G^0z)\ &=&\ 
[z^\mu (k_1)\, \Gamma^{HZG^0}_{0\mu}\, +\, \nonumber\\ 
&& z^\nu (k_2)\, \Gamma^{HG^0Z}_{0\nu}]\times\, \nonumber\\
&& \Delta_H (s)\
\bar{v}(p_2)\Gamma^{Hf\bar{f}}_0 u(p_1)\nonumber\\
{\cal T}^H_s  (zz) &=& z^\mu (k_1)z^\nu (k_2) 
{\cal T}^H_{s\,\mu\nu}(ZZ)\nonumber\\
&&{}
\eea   
with  $\Gamma^{HG^0G^0}_0   =    -i  g_w   M^2_H/(2M_W)$  and
$\Gamma^{HZG^0}_{0   \mu}    = -  g_w  (k_1  +   2  k_2)_\mu /(2c_w)$.
Evidently,  the  presence of  the term ${\cal  T}^H_P$ prevents ${\cal
  T}_s^H (Z_LZ_L)$  from satisfying the GET.    This is not surprising
however, since an  important Higgs-boson mediated $s$-channel part has
been  omitted.    Specifically,   the    momenta  $k_{1}^{\mu}$    and
$k_{2}^{\nu}$ stemming from  the  leading  parts of the   longitudinal
polarization        vectors          $\varepsilon^\mu_L(k_1)$      and
$\varepsilon^\nu_L(k_2)$    extract    such  a   term    from   ${\cal
  T}_t(Z_LZ_L)$.   Just as happens  in Eq.\  (\ref{TP}), this term  is
precisely  ${\cal T}^H_P$,   and  must be    added  to  ${\cal  T}_s^H
(Z_LZ_L)$, in  order  to form a  well-behaved  amplitude at very  high
energies.   In other  words,  the  amplitude $\widehat{{\cal   T}}^H_s
(Z_LZ_L)   = \\
{\cal T}^H_s (Z_LZ_L) +   {\cal T}^H_P$ satisfies the GET
independently ({\em cf.}\  Eq.\ (\ref{GETZZ})).  In fact, this crucial
property   persists after  resummation.  Indeed,   as  shown in  Fig.7
the  resummed amplitude $\overline{\cal T}^H_s (Z_LZ_L)$ may
be constructed from ${\cal  T}^H_s (Z_LZ_L)$ in Eq.\ (\ref{THsZZ}), if
$\Delta_H  (s)$ is   replaced by the  resummed Higgs-boson  propagator
$\hat{\Delta}^H  (s)$,    and $\Gamma^{HZZ}_{0  \mu\nu}$   by  the
expression $\Gamma^{HZZ}_{0\mu\nu} + \widehat{\Gamma}_{\mu\nu}^{HZZ}$,
where $\widehat{\Gamma}_{\mu\nu}^{ HZZ}$  is the one-loop $HZZ$ vertex
calculated  within the PT  \cite{PPHiggs}.  It is then straightforward
to  show that  the   Higgs-mediated amplitude $\widetilde{\cal  T}^H_s
(Z_LZ_L) = \overline{\cal T}^H_s (Z_LZ_L)\, +\, {\cal T}^H_P$ respects
the  GET {\em individually}; to that  end we only   need to employ the
following tree-level-type PT WI's:
\begin{eqnarray}
 \label{PTHZZ1}
&&k^\nu_2 \widehat{\Gamma}_{\mu\nu}^{HZZ}
(q,k_1,k_2) + iM_Z \widehat{\Gamma}^{HZG^0}_\mu (q,k_1,k_2)=\nonumber\\
&& -\, \frac{g_w}{2c_w}\, \widehat{\Pi}^{ZG^0}_\mu (k_1)\, ,\nonumber\\
  \label{PTHZZ2}
&&k^\mu_1 \widehat{\Gamma}_\mu^{HZG^0}
(q,k_1,k_2) + iM_Z \widehat{\Gamma}^{HG^0G^0}(q,k_1,k_2)=\nonumber\\
&& -\, \frac{g_w}{2c_w}\, \Big[\, \widehat{\Pi}^{HH}(q^2) + 
\widehat{\Pi}^{G^0G^0}(k^2_2)\, \Big]\, ,\nonumber\\
  \label{PTHZZ3}
&& k^\mu_1 k^\nu_2 \widehat{\Gamma}_{\mu\nu}^{HZZ}
(q,k_1,k_2) + M^2_Z \widehat{\Gamma}^{HG^0G^0}(q,k_1,k_2)=\nonumber\\
&& \frac{ig_wM_Z}{2c_w}\, \Big[\, \widehat{\Pi}^{HH}(q^2) + 
\widehat{\Pi}^{G^0G^0}(k^2_1) + \widehat{\Pi}^{G^0G^0}(k^2_2)\,
\Big]\, ,\nonumber\\
&&{}
\end{eqnarray}
where $\widehat{\Gamma}^{HZG^0}_\mu$  and $\widehat{\Gamma}^{HG^0G^0}$
are  the one-loop PT $HZG^0$ and  $HG^0G^0$ vertices, respectively. In
this derivation,  one should also make use  of the PT WI involving the
$ZG^0$- and  $G^0G^0$- self-energies:  $\widehat{\Pi}^{ZG^0}_\mu (k) =
-iM_Zk_\mu\, \widehat{\Pi}^{G^0G^0}(k^2)/k^2$.

The partial running widths for the Higgs boson 
have been first 
calculated at one-loop in \cite{PPHiggs}; they are given by:
\begin{eqnarray}
\Im m\ \widehat{\Pi}_{(WW)}(s) & =& 
\frac{\alpha_w}{16}\frac{M_H^4}{M_W^2}
\Big[\, 1+4\frac{M_W^2}{M_H^2}- \nonumber\\ 
&& 4\frac{M_W^2}{M_H^4}
(2s-3M_W^2)\, \Big]\beta_W \theta (s-4M_W^2)\, , \nonumber\\ 
\Im m\ {\widehat \Pi}_{(ZZ)}(s) &=& \frac{\alpha_w}{32}
\frac{M^4_H }{M^2_W}\Big[\, 1+4\frac{M^2_Z }{M^2_H} -\nonumber\\
&& 4\frac{M^2_Z }{M^4_H}(2s-3M^2_Z)\Big]
\beta_Z \theta (s-4M_Z^2)\, ,
\nonumber\\
\Im m\ {\widehat \Pi}_{(FF)}(s) &=& N_{F}
\frac{\alpha_w}{8}
\frac{m_{F}^2}{M^2_W}s\beta_{F}^3 
\theta (s-4m_{F}^2)\, ,\nonumber\\
\Im m\ {\widehat \Pi}_{(HH)}(s) &=& \frac{9\alpha_w}{32}\frac{M^4_H }{M^2_W}
\beta_H \theta (s-4M_H^2)\, .
\nonumber\\
&&{}
\label{widths}
\end{eqnarray}
In the above formula we denote by $F$ the various
fermionic flavours appearing inside the quantum loops,
i.e. $F \in \{ e,\mu ,\tau , u, d, c, s, t, b \}$.
$N_{F}=1$ for leptons, and $N_{F} =3$ for quarks.
In the case of a heavy Higgs boson the channels which dominate
numerically are the $WW$, $ZZ$ and $tt$. 

\section{Conclusions}

We have seen that, at one-loop order in perturbation theory,
 the Breit-Wigner resummation formalism can be
extended to the case of non-Abelian gauge theories, provided that
one resorts to the pinch technique rearrangement of the physical
amplitude. To accomplish this one needs invoke only the full
exploitation of the elementary Ward-identities of the theory,
in conjunction with unitarity, analyticity, and renormalization group
invariance. 

From   the phenomenological point  of  view the above
framework enables
the  construction  of Born-improved  amplitudes in  which   all
relevant physical information has been  encoded.  
This in turn will be very 
useful for the detailed study of the physical properties of particles,
most  importantly the correct  extraction of their masses, widths, and
line shapes. 

It would be very interesting to extend the formalism described in
this paper to higher orders. 
In addition, one would like to
reach 
a formal understanding
of the underlying mechanism, 
which at present can only be diagrammatically exposed. We  hope to be
able to return to these issues in the near future.

\section*{Acknowledgements} 
The research presented here has been carried out in collaboration
with Apostolos Pilaftsis, whom I thank for countless stimulating discussions
spanning a period of several years.
The contributions of Eduardo de Rafael and Jay Watson have 
been instrumental for my present understanding of the subject.
I also like to thank Fawzi Boudjema, 
Costas Kounnas, Kostas Philippides, Rob Pisarski, and
Raymond Stora, 
for various useful discussions.
This work has been
funded by  a Marie Curie  Fellowship (TMR-ERBFMBICT 972024).

\end{narrowtext}

\end{document}